\documentclass{webofc}
\usepackage[varg]{txfonts}   
\def\lc{L}
\def\lcp{{L}^{\prime}}
\def\lcpp{{L}^{\prime\prime}}
\def\pzp{p_0^{\prime}}
\def\pmp{p^{\prime}}
\def\ek{E_k}
\def\ep{E_p}
\def\eb{E_{\beta}}
\def\sqs{$\sqrt{s}$}

\begin{document}
\title{On the Relativistic Separable Functions for the Breakup\\ Reactions}
%
%

\author{\firstname{Serge G.} \lastname{Bondarenko}\inst{1}\fnsep\thanks{\email{bondarenko@jinr.ru}} \and
  \firstname{Valery V.} \lastname{Burov}\inst{1}
  \and
  \firstname{Elena P.} \lastname{Rogochaya}\inst{1}
}

\institute{Joint Institute for Nuclear Research, Dubna, Russia}

\abstract{%
In the paper the so-called modified Yamaguchi function
for the Bethe-Salpeter equation with a separable kernel is discussed.
The type of the functions is defined by the analytic stucture
of the hadron current with breakup - 
the reactions with interacting
nucleon-nucleon pair in the final state (electro-, photo-, and
nucleon-disintegration of the deuteron).
}
\maketitle
\section{Introduction}
\label{intro}
One of the most consistent nucleon-nucleon (NN) interaction theories
is based on the solution of the Bethe-Salpeter (BS) equation~\cite{Salpeter:1951sz}.
In this case, we have to deal with a nontrivial integral equation for the
bound state (deuteron) or interacting unbound NN pair.

The approximations based on the kernel with particle exchange are
hard to solve.
The one of the effective and solvable approach based on the exact solution of the BS
equation is to use the separable {\it Ansatz} for the interaction kernel
in the BS equation \cite{Bondarenko:2002zz}. In this case one can
transform an initial integral equation into a system of linear equations.
Parameters of the kernel are obtained by fitting of phase
shifts, inelasticity and low-energy parameters for respective partial-wave states.

First separable parametrizations were worked out within
nonrelativistic models. The separable functions (called form factors)
of the interaction kernel used in these models
had no poles on the real axis in the relative energy complex plane
\cite{Mathelitsch:1981mr,Haidenbauer:1984dz}.
However, such poles appeared when the interaction kernel was relativistically generalized.

In some cases they do not prevent to perform the calculations (for example
in elastic reactions).
However, at high energies, one have to deal with several
thresholds corresponding to the production of one, two and more
mesons of different types. Which is clearly not feasible to deal with. The more
practical approach is to employ a phenomenological covariant
separable kernel, which do not exhibit the meson-production
thresholds and can even be constructed in a singularity-free
fashion, using separable form factors and
Wick-rotation prescription as it is done in the present paper. Thus, an accurate description of
on-shell nucleon-nucleon data is possible up to quite high
energies. One then can hope that the obtained separable interaction kernels
have also a reasonable off-shell behavior, so that they can be applied to other reactions.

\section{Bethe-Salpeter formalism}
\label{sec-1}
We start with the partial-wave decomposed Bethe-Salpeter equation
for the nucleon-nucleon scattering matrix T (in the rest frame of two-nucleon system):
\begin{eqnarray}
t_{\lcp\lc}(\pzp, \pmp, p_0, p; s) &=&
v_{\lcp\lc}(\pzp, \pmp, p_0, p; s)
\label{t01}\\
&+& \frac{i}{4\pi^3} \sum_{\lcpp} \int\, dk_0\,\int\,
k^2\, dk\,
\frac{v_{\lcp\lcpp}(\pzp, \pmp, k_0, k; s)
t_{\lcpp\lc}(k_0, k, p_0, p; s)}{(\sqrt{s}/2-\ek+i\epsilon)^2-k_0^2}.
\nonumber\end{eqnarray}
Here $t$ is the partial-wave decomposed T matrix and
$v$ is the kernel of the $NN$ interaction, $\ek=\sqrt{k^2+m^2}$.
There is only one term in the sum for the singlet (uncoupled triplet) case
($L=J$) and there are two terms for the coupled triplet case ($L=J\mp1$).
We introduce square of the total momentum $s=P^2=(p_1+p_2)^2$ and the relative
momentum $p=(p_1-p_2)/2$ [$p^{\prime}=(p^{\prime}_1-p^{\prime}_2)/2$] (for details,
see reference~\cite{Bondarenko:2002zz}).

Assuming the separable form (rank I) for the partial-wave decomposed
kernels of NN interactions:
\begin{eqnarray}
v_{\lcp\lc}(\pzp, \pmp, p_0, p; s) = \lambda g^{[\lcp]}(\pzp, \pmp) g^{[\lc]}(p_0, p),
\label{t04}\end{eqnarray}
we can solve Eq.~(\ref{t01}) and write for the $T$ matrix:
\begin{eqnarray}
t_{\lcp\lc}(\pzp, \pmp, p_0, p; s) = \tau(s) g^{[\lcp]}(\pzp, \pmp) g^{[\lc]}(p_0, p),
\label{t05}\end{eqnarray}
with function $\tau(s)$ being:
\begin{eqnarray}
\tau(s) = 1/(\lambda^{-1} + h(s)).
\label{t06}\end{eqnarray}
Function $h(s)$ has the following form:
\begin{eqnarray}
h(s) = \sum_{\lc} h_{\lc}(s)= -\frac{i}{4\pi^3}\, \int\, dp_0\,\int\, p^2\, dp\,
\sum_{\lc} \frac{[g^{[\lc]}(p_0,p)]^2}{(\sqrt{s}/2-\ep+i\epsilon)^2-p_0^2}.
\label{t07}\end{eqnarray}

The simplest separable function $g(p_0,p)$ which can be used, is a covariant generalization of the non-relativistic
{\em Yamaguchi}-type~\cite{yam,yamb} function:
\begin{eqnarray}
g(p_0,p) = \frac{1}{p_0^2-p^2-\beta^2+i\epsilon},
\label{t11a}\end{eqnarray}
where $\beta$ is a parameter.

\subsection{Modified Yamaguchi-type functions}
\label{sec-2}

Let us consider the integral $h(s)$ (Eq.~\ref{t07}). Taking into account the
pole structure of the propagators:
\begin{eqnarray}
p^{(1,2)}_0=\pm \sqrt s/2 \mp \ep\pm i\epsilon
\label{t12a}\end{eqnarray}
and of $g$ functions:
\begin{eqnarray}
p^{(3,4)}_0=\mp \eb \pm i\epsilon
\label{t12b}\end{eqnarray}
and using the Cauchy theorem, the $h(s)$ function can be written as follows:
\begin{eqnarray}
\frac{1}{2\pi^2}\int p^2 dp \frac{1}{(s/4-\sqrt s \ep+m^2-\beta^2)^2} \frac{1}{\sqrt s-2\ep+i\epsilon}.
\label{t13}\end{eqnarray}

To calculate the integral Eq.~(\ref{t13}) one should analyze the numerator
$f = (s/4-\sqrt s \ep+m^2-\beta^2)$ as a function of $s$:
\begin{itemize}
\item
if $2(m-\beta)<$ \sqs $<2(m+\beta)$ then always $f < 0$ and the function ${1}/{f^n}$
{\it is integrable} for any integer $n$ and any $\ep$;
\item
for a bound state \sqs $=M_d=(2m-\epsilon_D)$. Since for minimal $\beta_{\min}=0.2$ GeV
always $\beta_{\min} > \epsilon_D/2$ then the function ${1}/{f^n}$ {\it is integrable}
for any integer $n$ and any $\ep$;
\item
if \sqs $<2(m-\beta)$ or \sqs $>2(m+\beta)$ then $f$ can be positive and negative
and ${1}/{f^n}$ {\it is non-integrable} for even $n$ at any $\ep$.
\end{itemize}

The critical value $s^{c}=4(m+\beta)^2$ corresponds
to the laboratory kinetic energy of $np$-pair $T_{lab}^c = 4\beta + 2\beta^2/m
\simeq 4\beta$. If $\beta_{\min}=0.2$ GeV then $T_{lab}^{\min} = 0.8$ GeV.

So, if we consider breakup processes of the deuteron such as
photo-, electro- and nucleon-breakup Yamaguchi-functions can be used only if the
laboratory kinetic energy of the NN-pair is less than $T_{lab}^{\min}$.
To avoid this restriction we suggest to use Yamaguchi-type functions modified
in the following way:
$$
g_{\rm Y}(p_0,p) = 1/(p_0^2-p^2-\beta^2) \longrightarrow g_{\rm MY}(p_0,p) =
1/((p_0^2-p^2-\beta^2)^2+\alpha^4),
$$
here Y stands for the Yamaguchi and MY -- for Modified Yamaguchi functions.

To work with the modified Yamaguchi-type functions the procedure of $p_0$ integration
should be modified, too. This procedure is worthy of a special discussion.
The poles of the $h(s)$ integral with the modified Yamaguchi-type functions are:
\begin{eqnarray}
p^{(3,4)}_0 = \pm \sqrt{p^2+\beta^2 + i\alpha^2},
\nonumber\\
p^{(5,6)}_0 = \pm \sqrt{p^2+\beta^2 - i\alpha^2}. \label{p0a2}
\end{eqnarray}
%
\begin{figure}[ht]
\begin{center}
\includegraphics[width=0.45\textwidth]{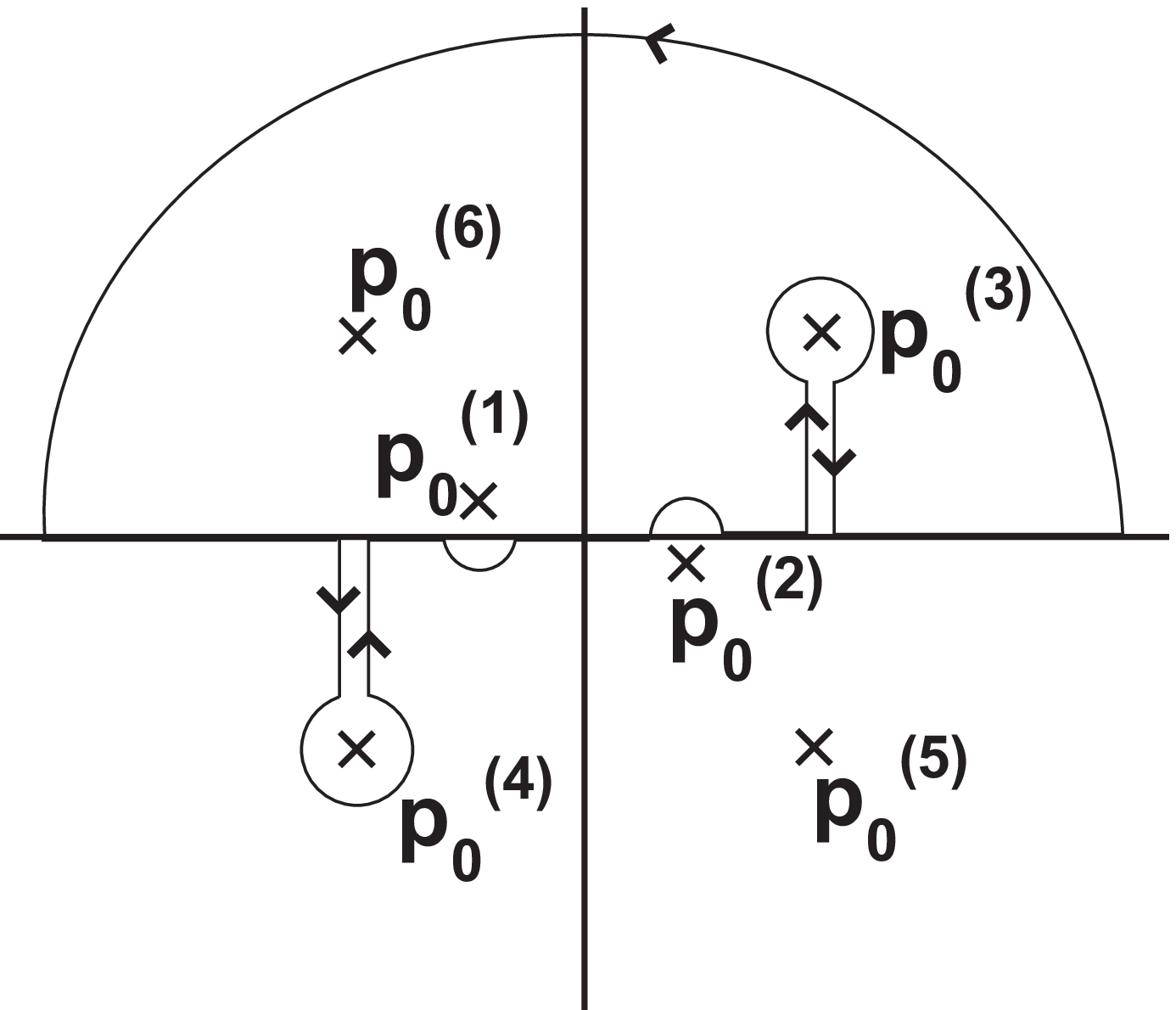}
\includegraphics[width=0.45\textwidth]{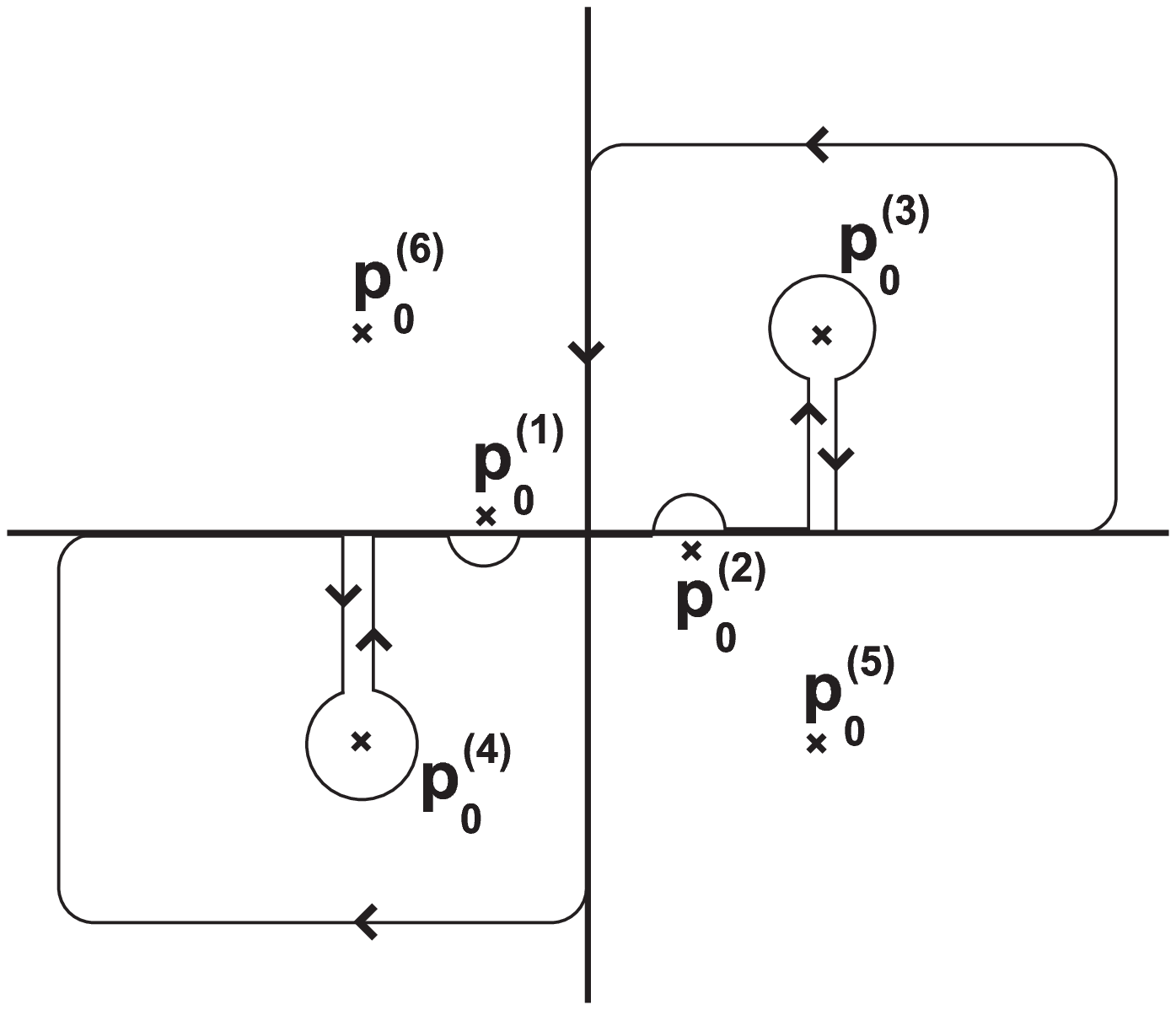}
\caption{
  (a) Contour for integration over $p_0$ according to the Cauchy theorem.
  (b) Contour for integration over $p_0$: the Wick rotation.
}
\label{contour1}
\end{center}
\end{figure}
%

All poles and the contour of integration
are pictured in Fig.~\ref{contour1}(a,b). The idea how to choose the
contour appeared owing to \cite{Cutkosky:1969fq,Lee:1969fy}. It
is:
\begin{enumerate}
\item the contour must envelope the poles of $g$ form
factors which will be inside the standard contour in the
$\alpha\rightarrow 0$ limit. ``Standard'' means the contour used in the
quantum field theory calculations with a propagator which has
poles only on the real axis in the $p_0$ complex plane; one of
them is circled from below and the other -- from above. So, the path
of integration is defined by an appropriate contour for the
propagator
\item the calculation over the presented path leads to the
pure real contribution from the form factor poles and, therefore,
to the unitary $S$ matrix (or the corresponding unitarity condition for the T matrix).
We also obtain a correct transition to
ordinary form factors of type $g\sim 1/(p_0^2-p^2-\beta^2)^2$ in the
$\alpha\rightarrow 0$ limit.
\end{enumerate}

In general, the modified Yamaguchi-type functions can be written as:
\begin{eqnarray}
g_i^{[a]}(p_0,p) = \frac{(p_{ci}-p_0^2+p^2)^{n_i}(p_0^2-p^2)^{m_i}}
{((p_0^2-p^2-\beta_{1i}^2)^2+\alpha_{1i}^4)^{k_i}
((p_0^2-p^2-\beta_{1i}^2)^2+\alpha_{1i}^4)^{l_i}},
\label{t15}\end{eqnarray}
where parameters -- $n_i,m_i,k_i,l_i$ (integer), $p_{ci},\beta_{1i},\beta_{2i},\alpha_{1i},\alpha_{2i}$
(real) depend on the channel $[a]$ under consideration. Such $g$ form factors are used to describe
neutron-proton scattering observables (phase shifts, inelasticities, low-energy
parameters and deuteron characteristics) for the total angular momentum $J=0,3$ in a wide
energy range (see~\cite{Bondarenko:2008mm}--\cite{Bondarenko:2014}).

\section*{Acknowledgement}
One of the authors (S.G. Bondarenko) would like to thank the organizers
of the International Conference ``Mathematical Modeling and Computational Physics, 2017''
(JINR, Dubna, July 3--7, 2017).
for the invitation and opportunity to present our results.

This work was partially supported by the Russian Foundation for Basic Research
grant $N\textsuperscript{\underline{o}}$ 16-02-00898.


\begin{thebibliography}{10}
%
%
\bibitem{Salpeter:1951sz} E.E.~Salpeter and H.A.~Bethe, Phys. Rev. {\bf 84}, 1232 (1951)

\bibitem{Bondarenko:2002zz} S.G.~Bondarenko, V.V.~Burov, A.V.~Molochkov,
G.I.~Smirnov, and H.~Toki, Prog. Part. Nucl. Phys. {\bf 48}, 449
(2002)

\bibitem{Mathelitsch:1981mr} L.~Mathelitsch, W.~Plessas, and W.~Schweiger,
Phys. Rev. {\bf C26}, 65 (1982)

\bibitem{Haidenbauer:1984dz}
J.~Haidenbauer and W.~Plessas, Phys. Rev. {\bf C30}, 1822 (1984)

\bibitem{yam} Y.~Yamaguchi, Phys. Rev. {\bf 95}, 1628 (1954)
\bibitem{yamb}Y.~Yamaguchi, Y.~Yamaguchi, Phys. Rev. {\bf 95}, 1635 (1954)

\bibitem{Cutkosky:1969fq}
R.E.~Cutkosky, P.V.~Landshoff, D.I.~Olive, and J.C.~Polkinghorne, Nucl. Phys. {\bf B12}, 281 (1969)

\bibitem{Lee:1969fy}
T.D. Lee and G.C. Wick, Nucl. Phys. {\bf B9}, 209 (1969)

\bibitem{Bondarenko:2008mm}
S.G. Bondarenko, V.V. Burov, W.-Y. Pauchy Hwang, and E.P.
Rogochaya, Nucl. Phys. {\bf A832}, 233 (2010) 

\bibitem{Bondarenko:2010}
S.G. Bondarenko, V.V. Burov, W.-Y.~Pauchy Hwang, and E.P. Rogochaya,
Nucl. Phys. {\bf A848}, 75 (2010) 

\bibitem{Bondarenko:2011hs}
S.G. Bondarenko, V.V. Burov, and E.P. Rogochaya,
Phys. Lett. {\bf B705}, 264-268 (2011);
Nucl. Phys. Proc. Suppl. {\bf 219--220}, 126--129 (2011)

\bibitem{Bondarenko:2013}
S.G. Bondarenko, V.V. Burov, and E.P. Rogochaya,
Nucl. Phys. Proc. Suppl. {\bf 245}, 291--297 (2013)

\bibitem{Bondarenko:2014}
S.G. Bondarenko, V.V. Burov, S.E. Kemelzhanova, E.P. Rogochaya, and N. Sagimbaeva,
\emph{Proceedings of the 12th International Workshop ``Relativistic Nuclear Physics from
Hundreds of MeV to TeV''} (Slovak Republic, Stara Lesna, June 16--20, 2014)
\end{thebibliography}
\end{document}